\documentclass[apj,numberedappendix,epsfigure]{emulateapj}
\usepackage{apjfonts}

\usepackage[normalem]{ulem}

\shorttitle{SEARCH FOR PULSATING HE WHITE DWARFS}
\shortauthors{STEINFADT ET AL.}

\newcommand{\be}{\begin{eqnarray}}
\newcommand{\ee}{\end{eqnarray}}

\newcommand{\phz}{\phantom{0}}

\begin{document}


\title{A Search for Pulsations in Helium White Dwarfs}

\author{Justin D. R. Steinfadt\altaffilmark{1}}

\author{Lars Bildsten\altaffilmark{1,2}}

\author{David L. Kaplan\altaffilmark{3}}

\author{Benjamin J. Fulton\altaffilmark{4}}

\author{Steve B. Howell\altaffilmark{5}}

\author{T. R. Marsh\altaffilmark{6}}

\author{Eran O. Ofek\altaffilmark{7,8}}

\author{Avi Shporer\altaffilmark{1,4}}

\altaffiltext{1}{Department of Physics, Broida Hall, University of California,
Santa Barbara, CA 93106; jdrsteinfadt@gmail.com}
\altaffiltext{2}{Kavli Institute for Theoretical Physics and Department of Physics, Kohn Hall, University of California, Santa Barbara, CA 93106; bildsten@kitp.ucsb.edu}
\altaffiltext{3}{Physics Department, University of Wisconsin - Milwaukee, Milwaukee WI 53211; kaplan@uwm.edu}
\altaffiltext{4}{Las Cumbres Observatory Global Telescope, 6740 Cortona Drive, Suite 102, Goleta, CA 93117; bjfulton@lcogt.net, ashporer@lcogt.net}
\altaffiltext{5}{NASA Ames Research Center}
\altaffiltext{6}{Department of Physics, University of Warwick, Coventry CV4 7AL, UK}
\altaffiltext{7}{Division of Physics, Mathematics and Astronomy, California Institute of Technology, Pasadena, CA 91125}
\altaffiltext{8}{Einstein Fellow}


\begin{abstract}

The recent plethora of sky surveys, especially the Sloan Digital Sky Survey, have discovered many low-mass ($M<0.45M_\odot$) white dwarfs that should have cores made of nearly pure helium. These WDs come in two varieties; those with masses $0.2<M<0.45 M_\odot$ and H envelopes so thin that they rapidly cool, and those with $M<0.2M_\odot$ (often called extremely low mass, ELM, WDs) that have thick enough H envelopes to sustain $10^9$ years of H burning. In both cases, these WDs evolve through the ZZ Ceti instability strip, $T_{\rm eff}\approx 9$,000--12,000\,K, where $g$-mode pulsations always occur in Carbon/Oxygen WDs. This expectation, plus theoretical work on the contrasts between C/O and He core WDs, motivated our search for pulsations in 12 well characterized helium WDs. We report here on our failure to find any pulsators amongst our sample. Though we have varying amplitude limits, it appears likely that the theoretical expectations regarding the onset of pulsations in these objects requires closer consideration.  We close by encouraging additional observations as new He WD samples become available, and speculate on where theoretical work may be needed.

\end{abstract}

\keywords{stars: white dwarfs--- stars: oscillations}


\section{Introduction}

White dwarf asteroseismology offers the opportunity of directly constraining mass, core composition, envelope composition and stratification, spin rate, and magnetic field strength \citep{cas08.mnras385}.  The ZZ~Ceti H-atmosphere white dwarfs (WDs) pulsate in normal modes of oscillation ($g$-modes) where buoyancy is the restoring force \citep{win08.araa46}.  These stars exhibit pulsations when they enter the instability region, a discrete strip in the $T_{\rm eff}$--$\log g$ plane that spans 11,000 $\lesssim T_{\rm eff} \lesssim$ 12,250\,K at $\log g = 8.0$.  This strip has been extensively observed and constrained by numerous sources \citep{wes91.pro91,muk04.apj607,cas07.aap462,gia07.pro372} and is likely pure; i.e., all  cooling WDs pulsate as they evolve through the instability strip.  These data offer insight into the evolution of WDs \citep{fon82.apj258,cas07.apj661}.  The lowest mass WD known to pulsate is HE~0031-5525 at $\log g = 7.65$ and $T_{\rm eff} = 11,480$\,K and a mass of $\approx$0.5$M_{\odot}$ \citep{cas06.aap450}.  The known ZZ~Ceti pulsators are certainly C/O-core WDs, where the theoretical understanding has been focused \citep{win82.apj252,bra91.apj267,bra97.pro214,win08.araa46}.

Starting with \citet{ber92.apj394} and \citet{mar95.mnras275}, there has been a growing realization that many WDs have $\log g \lesssim 7.76$, implying core masses lower than expected for C/O-core WDs.  These are expected to be He WDs made from the truncated red giant branch (RGB) evolution of $M<2.3$\,$M_{\odot}$ stars \citep{dcr96.apj466,dom99.apj524,pie04.apj612,van06.apjs162,pan07.mnras382}.  To form a He WD, a process removes mass from the H envelope, quenching the H-shell burning that grows the He core to ignition.  Two modes of evolution have been proposed to truncate RGB evolution: mass loss due to binary interaction and mass loss due to stellar winds.  As a progenitor star ascends the RGB, its radius increases rapidly, eventually filling the Roche Lobe, and mass transfer begins.  If unstable, this mass transfer initiates a common envelope event \citep{pac76.pro73,ibe93.pasp105} ejecting the envelope from the red giant as the He-core and companion spiral into shorter orbital periods \citep{taa00.araa38,del10.apj719,ge10.apj717}.  It is this physics that leads to the expectation that He WDs should be in binary systems \citep{mar95.mnras275,kil07.apj671,agu09.apj697}.  However, surveys have uncovered significant numbers of He WDs that do not appear to be in binary systems \citep{agu09.apj697,bro10.apj723,bro11.apj730}.  In stars of high metallicity, the increased opacity in the envelope due to metals can drive very strong stellar winds that cause mass loss in excess of 0.5$M_{\odot}$ during the RGB phase \citep{han94.mnras270,lee94.apj423,cat00.apj531,han05.apj635}.  Therefore, it is possible that high metallicity stars may lose enough mass on the RGB due to stellar winds to avoid He core ignition and thus leave $M\gtrsim0.4$\,$M_{\odot}$ He WDs as remnants.  Direct evidence for this scenario is scarce, but may exist in the over-luminous low mass He WDs in the globular cluster NGC~6791 \citep{han05.apj635,kal07.apj671,gar10.nat465}.  It is clear that this wind mass loss scenario cannot produce low mass He WDs.

\begin{figure*}
	\centering
	\includegraphics{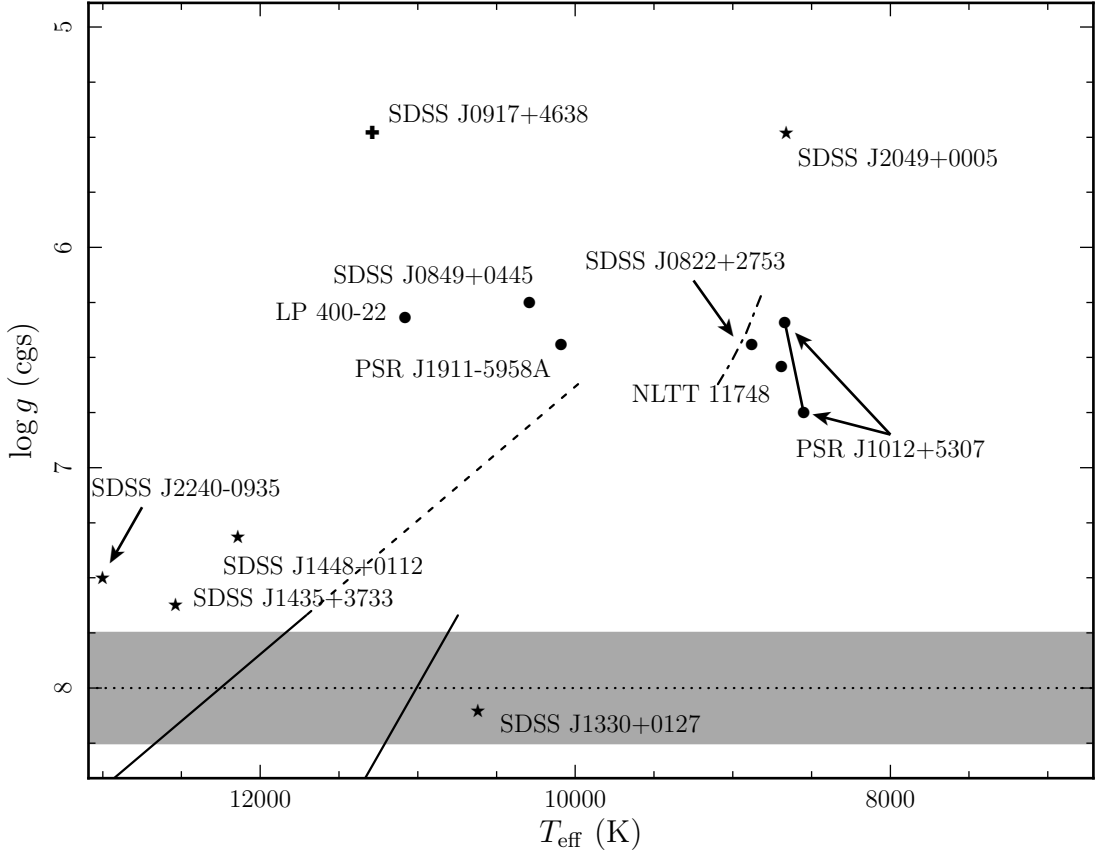}
	\caption{ \small The $\log g$-$T_{\rm eff}$ plane.  The dark-grey horizontal strip illustrates the location of the C/O WDs.  The solid lines represent the fitted boundaries of the empirical ZZ~Ceti instability strip as determined by \citet{gia07.pro372} with the dashed line merely being the extension of the blue edge.  The dash-dotted line represents the theoretical blue edge highlighted for very low mass He WDs by \citet{ste10.apj718}.  We plot only objects we have observed.  SDSS~J1330+0127, SDSS~J1435+3733, SDSS~J1448+0112, SDSS~J2049+0005, and SDSS~J2240-0935, the stars, are from \citet{eis06.apjs167} (corrected by \citealt{sil06.aj131} and \citealt{hel09.aap496}).  SDSS~J0822+2753 and SDSS~J0849+0445 are from \citet{kil10.apj716}.  SDSS~J0917+4638, the plus, is from \citet{kil07.apj660}.  NLTT~11748 is from \citet{kaw09.aap506}, \citet{ste10.apj716}, and \citet{kil10.apj721}.  LP~400-22 is from \citet{kaw06.apj643}.  PSR~J1012+5307 is from two sources, neither obviously superior, \citet{van96.apj467} and \citet{cal98.mnras298}.  PSR~J1911-5958A is from \citet{bas06.aap456}.}
	\label{fig:obscand}
\end{figure*}

Once formed, He WD evolution is different from that of C/O WDs because nuclear burning may still play a significant role, especially when $M\lesssim0.2$\,$M_{\odot}$.  Studies of millisecond pulsars suggest that many He WDs are born with thick $\sim$10$^{-3}$--10$^{-2}M_{\odot}$ H envelopes \citep{alb96.nat380,van05.pro328,bas06.aap456,bas06.aap450}.  These thick H envelopes develop long diffusive tails into the He core where the higher densities and temperatures can cause them to ignite \citep{dri99.aap350,ser02.mnras337,pan07.mnras382}.  This ignition highlights a dichotomy in the evolution of He WDs.  For He WDs of $\gtrsim$0.18--0.20$M_{\odot}$ (dependent upon metallicity) the H envelopes are radially thin enough that thin-shell instabilities develop and a series of rapid thermonuclear flashes occur.  These flashes burn away the envelope leaving only a thin layer of H which allows for rapid cooling on timescales of 10--100s of Myrs.  For He WDs of $\lesssim$0.18--0.20$M_{\odot}$, the nuclear burning is stable and provides a dominant energy source that slows the cooling evolution of the WD to timescales of several Gyrs \citep{ser02.mnras337,pan07.mnras382}.  From an observational standpoint, very low mass He WDs ($<$0.2$M_{\odot}$) may spend significantly more time in a ZZ~Ceti like instability strip than the higher mass He WDs.  However, no systematic search for their pulsations has been undertaken.  This is the goal of our work.

Some differences from the C/O-core ZZ~Ceti stars are expected.  \citet{ste10.apj716} used very low mass He WD models ($<$0.2$M_{\odot}$) with stable H burning envelopes and the Brickhill convective driving criterion \citep{bri83.mnras204,bri91.mnras251,wu99.apj519} to locate He WD pulsation candidates.  The result of this work is seen in the ``blue" edge boundary depicted between $\log g = $6.0--6.5 in Figure \ref{fig:obscand}. Clearly, many theoretical issues remain in pinpointing the precise location of the blue edge, such as the effects of convective efficiency and the use of more detailed fully non-adiabatic calculations, which may move this boundary by several hundred Kelvin \citep{win82.apj262,fon08.pasp120}.  For normal mass He WDs (0.2--0.45\,$M_{\odot}$), the blue edge of the instability strip has not been theoretically determined (though, see \citealt{arr06.apj643} from some preliminary calculations) and so we naively extrapolate the empirical boundary of the ZZ~Ceti stars as determined by \citet{gia07.pro372} and connect it to the very low mass He WD boundary. Given the many theoretical uncertainties and naive extrapolations, the boundary depicted in Figure \ref{fig:obscand} is meant only to highlight the parameter space in which to carry out our search.

In this paper we present our search for He WD pulsators.  In Section \ref{sec:find}, we describe our selection criteria for observing candidate pulsators using objects culled from a multitude of sources.  In Section \ref{sec:obsstrat}, we describe our observational strategy and analysis procedures.  In Section \ref{sec:obssum}, we present our observations of 12 candidate objects and our detection limits.  No significant pulsations have been detected, however, our detection limits may provide interesting constraints on theory as we describe in Section \ref{sec:conc}.

\begin{deluxetable*}{l c l l c c}
\centering
\tabletypesize{\footnotesize}

\setlength{\tabcolsep}{6pt}
\tablewidth{0pt}
\tablecaption{ Observed Candidates and Results\label{tbl:candsum}}

\tablehead{
\multicolumn{3}{l}{\hspace{ 9.5 mm }Object\hspace{ 9.25 mm }$g^\prime$-SDSS\hspace{ 5.5 mm }$\log g$} & \colhead{$T_{\rm eff}$} & \multicolumn{2}{l}{\hspace{ 12 mm }Reference\hspace{ 5.25 mm }Det. Limit\tablenotemark{a}} \\
\multicolumn{3}{l}{\hspace{ 31.0 mm }(mag)\hspace{ 6.75 mm }(cgs)} & \colhead{(K)} & \multicolumn{2}{l}{\hspace{ 36.5 mm }(mmag)}
 }
\tablecolumns{6}

\startdata

	\object[SDSS J082212.57+275307.4]{SDSS~J0822+2743} & 18.3 & $6.55\pm0.11$ & $\phz8,880\pm60$ & \citet{kil10.apj716} & 8  \\
	\object[SDSS J084910.13+044528.7]{SDSS~J0849+0445} & 19.3 & $6.23\pm0.08$ & $10,290\pm250$ & \citet{kil10.apj716} & 15 \\
	\object[SDSS J091709.55+463821.8]{SDSS~J0917+4638} & 18.7 & $5.48\pm0.03$ & $11,288\pm72$ & \citet{kil07.apj660} & 11 \\
	\object[SDSS J133058.19+012706.5]{SDSS~J1330+0127} & 18.9 & $8.1\phz\pm0.2$ & $10,617\pm452$ & \citet{sil06.aj131} & 44 \\
	\object[SDSS J143547.87+373338.5]{SDSS~J1435+3733} & 17.1 & $7.62\pm0.12$ & $12,536\pm488$ & \citet{eis06.apjs167} & 4 \\
	\object[SDSS J144859.39+011243.7]{SDSS~J1448+0112} & 19.6 & $7.31\pm0.18$ & $12,142\pm486$ & \citet{eis06.apjs167} & 34 \\
	\object[SDSS J204949.78+000547.3]{SDSS~J2049+0005} & 19.7 & $5.48\pm0.10$ & $\phz8,660\pm144$ & \citet{eis06.apjs167}\tablenotemark{b} & 21 \\
	\object[SDSS J224038.38-093541.3]{SDSS~J2240-0935} & 17.6 & $6.96\pm0.09$ & $11,449\pm205$ & \citet{hel09.aap496} & 14 \\
	\object[LP 400-22]{LP~400-22} & 17.2 & $6.32\pm0.08$ & $11,080\pm140$ & \citet{kaw06.apj643} &  4 \\
	\object[NLTT 11748]{NLTT~11748} & 17.1\tablenotemark{c} & $6.54\pm0.05$ & $\phz8,690\pm140$ & \citet{kil10.apj721} & 5 \\
	\object[PSR J1012+5307]{PSR~J1012+5307} & 19.5\tablenotemark{d} & $6.34\pm0.20$ & $\phz8,670\pm300$ & \citet{cal98.mnras298} & 20 \\
	   &   & $6.75\pm0.07$ & $\phz8,550\pm25$ & \citet{van96.apj467} & \\
	\object[PSR J1911-5958A]{PSR~J1911-5958A} & 22.2\tablenotemark{d} & $6.44\pm0.20$ & $10,090\pm150$ & \citet{bas06.aap456} & 16 \\
\enddata
\tablenotetext{a}{Reporting the best limit over a minimum of 1--7\,mHz frequency range (where possible).  See Table \ref{tbl:results} for frequency range details and multiple epochs.}
\tablenotetext{b}{SDSS~J2049+0005 is likely a distant A star \citep{eis06.apjs167,kil07.apj660}.}
\tablenotetext{c}{Magnitude in Johnson $B$.}
\tablenotetext{d}{Magnitude in Johnson $V$.}
\end{deluxetable*}


\section{ Determining the He WD Targets }
\label{sec:find}

To select He WD candidates for observation we generally select only those WDs with $\log g < 7.6$ (a mass $<$0.45\,$M_{\odot}$) and $7,000 < T_{\rm eff} < 12,000$\,K.  We select a larger range of temperatures than suggested by ZZ~Ceti properties (and the theoretical analysis of \citealt{ste10.apj718}), as non-detections will constrain the parameter space in which the conventional theory must hold.  Of added interest is the potential detection of pulsations in parameter space not predicted by theory.

We find most of our candidates from surveys looking for other phenomena, e.g. the Sloan Digital Sky Survey (SDSS), but we have also found candidates in single object papers.  Companions to pulsars have been found to be good candidates such as PSR~J1911-5958A \citep{bas06.aap456} and PSR~J1012+5307 \citep{van96.apj467,cal98.mnras298}.

\citet{eis06.apjs167} offers a catalog of WDs within the SDSS that is the largest to date.  They found their WDs by selecting SDSS spectra using a series of color cuts and then fitting to a large grid of WD atmospheric models.  Objects with spectral contamination due to a companion star, e.g. an M-dwarf star, were not de-contaminated and thus the spectral fits obtained were often erroneous.  This is an issue for He WDs as their primary formation channel requires a binary companion which frequently contaminates the spectrum.  Many of these contaminated spectra were reanalyzed by \citet{sil06.aj131} and \citet{hel09.aap496} in two catalogs of close binary systems.  In these cases, the derived parameters of \citet{sil06.aj131} and \citet{hel09.aap496} were used.  Since the SDSS sample did not target WDs specifically, the quality of the spectra of many of the WDs is low as is the model fits derived from them.  Therefore, we generally do not pursue SDSS targets unless additional spectra have been taken.  A few SDSS objects were observed early in our program before this contamination was well understood.

\citet{bro06.apj647} performed a hypervelocity star survey looking for B-type stars escaping the galaxy.  They used SDSS photometry to color select their B-type stars and gathered high signal-to-noise spectra to search for large radial velocities.  Given their color selection, \citet{kil07.apj660,kil07.apj671,kil10.apj716} made great use of this data-set, uncovering many low mass He WDs.  \citet{kaw06.apj643} and \citet{kaw09.aap506} have used a similar technique to the \citet{bro06.apj647} survey and searched for hypervelocity stars in proper motion data from a variety of sources (\citealt{osw93.pro403}, New Luyten Two Tenths, NLTT, Proper Motion Survey).  Their survey has discovered some of the lowest mass WDs known \citep{kaw06.apj643,kaw09.aap506}.  These surveys revealed dozens of candidates, only 12 of which are observed; see Table \ref{tbl:candsum}.  Of these 12 objects, 10 are potential He WDs and 7 are very low mass He WDs.

\section{ Observational Strategy }
\label{sec:obsstrat}

The observations we executed followed a simple strategy.  \citet{ste10.apj718} showed that the expected periods of pulsations range between 250 and 1,000 seconds.  Extrapolating our knowledge of the ZZ~Ceti stars we considered a fiducial amplitude of 10 mmags.  We designed each observation to be sensitive to these properties by incorporating a fast cadence (usually less than 60 seconds) and individual measurements of the target that met or exceeded 10 mmags in precision (in the Poisson limit, 10,000 photons/image).  In Table \ref{tbl:detobs} we report on the details of our observations.

\begin{deluxetable*}{lclcccc}
\tabletypesize{\footnotesize}

\setlength{\tabcolsep}{5pt}
\tablewidth{0pt}
\tablecaption{Details of Observations\label{tbl:detobs}}

\tablehead{
\colhead{Object} & \colhead{$g^\prime$-SDSS} & \colhead{Telescope/} & \colhead{Date} & \colhead{Len.} & \colhead{Exp.} & \colhead{N. of} \\
	& \colhead{(mag)} & \colhead{Camera/Filter} & \colhead{\fontsize{8}{8}\selectfont (yyyymmdd)} & \colhead{(hrs.)} & \colhead{Time (s)} & \colhead{Exp.\tablenotemark{a}}
 }
\tablecolumns{7}

\startdata

	SDSS~J0822+2743 & 18.3 & WHT/ACAM/$g^\prime$ & 20110220 & 2.1 & 30 & 137 \\
	SDSS~J0849+0445 & 19.3 & WHT/ACAM/$g^\prime$ & 20110117 & 1.9 & 60 & 96 \\
	SDSS~J0917+4638 & 18.7 & WHT/ACAM/$g^\prime$ & 20110220 & 2.3 & 45 & 145 \\
	SDSS~J1330+0127 & 18.9 & WIYN/OPT/BG39 & 20070530 & 1.6 & 90 & 57 \\
	SDSS~J1435+3733 & 17.1 & WIYN/OPT/BG39 & 20070529 & 1.5 & 15 & 197\tablenotemark{b} \\
	                                      & & WIYN/OPT/BG39 & 20070531 & 1.9 & 15 & 233\tablenotemark{b} \\ 
	                                      & & WIYN/OPT/BG39 & 20070601 & 3.8 & 45 & 259\tablenotemark{b} \\
	SDSS~J1448+0112 & 19.6 & WIYN/OPT/BG39 & 20070531 & 2.0 & 60 & 104 \\
	SDSS~J2049+0005 & 19.7 & WIYN/OPT/BG39 & 20070531 & 3.4 & 60 & 166 \\
	SDSS~J2240-0935 & 17.6 & P60/P60CCD/$g$ & 20060804 & 3.7 & 30 & 204 \\
	                                     & & P60/P60CCD/$g$ & 20060817 & 4.5 & 100 & 73\tablenotemark{c} \\
	                                     & & P60/P60CCD/$g$ & 20060919 & 1.1 & 30 & 57 \\
	                                     & & P60/P60CCD/$g$ & 20060925 & 1.0 & 30 & 60 \\
	                                     & & P60/P60CCD/$g$ & 20061205 & 0.9 & 30 & 59 \\
	LP~400-22 & 17.2 & WIYN/OPTIC/$g^\prime$ & 20070530 & 2.4 & 20 & 269 \\
	NLTT~11748 & 17.1\tablenotemark{d} & FTN/Merope/$g^\prime$ & 20091222 & 4.0 & 45 & 198\tablenotemark{b} \\
	                         & & FTN/Merope/$g^\prime$ & 20100108 & 3.3 & 45 & 167\tablenotemark{b} \\
	PSR~J1012+5307 & 19.5\tablenotemark{e} & WHT/ACAM/$g^\prime$ & 20100330 & 2.0 & 60 & 103 \\
	PSR~J1911-5958A & 22.2\tablenotemark{d} & HST/WFC3/F200LP & 20100305 & 6.0 & 60 & 142 \\
\enddata
\tablenotetext{a}{Excludes images contaminated by cosmic rays, clouds, and excess noise.}
\tablenotetext{b}{Also excludes images taken during eclipse.}
\tablenotetext{c}{Observation contains large gaps.}
\tablenotetext{d}{Magnitude in Johnson $B$.}
\tablenotetext{e}{Magnitude in Johnson $V$.}
\end{deluxetable*}

\subsection{ Aperture Photometry }

All images were reduced using standard methods (bias subtracted and flat fielded).  In most cases, the CCD detectors under-sampled the point spread function (PSF) of the stars making PSF-fitting photometry highly imprecise.  Therefore, aperture photometry was executed using the Image Reduction and Analysis Facility (IRAF)\footnote{IRAF is distributed by the National Optical Astronomy Observatory, which is operated by the Association of Universities for Research in Astronomy, Inc., under contract with the National Science Foundation.  http://iraf.noao.edu} software package {\it apphot}.  The only exception to this analysis is PSR~J1911-5958A where unique HST observations required a different approach which will be discussed later.

We performed differential photometry using multiple comparison stars inversely weighted to their intrinsic variance over a single run of observation (a technique similar to that used in \citealt{gil88.pasp100} and \citealt{sok01.mnras326}).  We ensure all comparison stars are non-variable by comparing each prospective star within the field-of-view with all others in every combination and use periodograms to ensure no periodic trends are present.

We employ a variable aperture algorithm to optimize the signal-to-noise of our photometry in order to deal with the varying PSF.  For each observation, a single representative frame was selected.  For every star in the image we varied the size of the photometric aperture to find that size which optimized the signal-to-noise of the measurement.  The optimum apertures were then scaled by the average FWHM of 2D-gaussians fitted to each star.  For all other images the average FWHM of the stellar PSFs was measured and then scaled using the appropriate optimum scaling.  Therefore, as the seeing fluctuates during an observation the apertures are adjusted larger or smaller (a technique similar to that used in \citealt{dee01.pro01}).

Once we have constructed the light curve for our target star there are often long timescale trends still present in the data, usually caused by the color-airmass effect.  We correct for this effect by correlating our light curve with airmass and dividing a fitted trend from the light curve.  For some observations this trend removal is insufficient as other systematics may be present (e.g. variable sky background due to the moon).  In these cases low order polynomials (third order or less) are used to remove this power.

\subsection{ Periodograms and Detection Limits }

To detect pulsations in our light curve, we use a Lomb-Scargle periodogram \citep{sca82.apj263} because our observations are frequently unevenly spaced and sometimes filled with large gaps.  In the ZZ~Ceti stars, low amplitude pulsations show pulse shapes that are closely approximated by sinusoids while high amplitude pulsations frequently show narrow peaks in their shape \citep{fon08.pasp120}.  We feel this justifies our use of a periodogram which is optimized for sinusoidal signals because the pulsations that are ``in the noise" are closely represented by sinusoids.  High amplitude pulsations, though non-sinusoidal, should still be easily detected as they will likely be well above the noise.

We have not found any significant pulsations from He WDs.  However, detection limits are still useful in both deciding if additional observations are required and constraining theory.  We measure our detection limits by using signal injection and the properties of a Lomb-Scargle periodogram.  For a periodogram normalized by the variance of the total light curve, a false-alarm probability is naturally generated that states for $M$ independent frequency bins the probability that independent and normal noise produces power greater than $z$ in any of the frequency bins is ${\rm Prob(Any\,Power} > z) = 1 - (1 - e^{-z})^M$ \citep{sca82.apj263,hor86.apj302}.  Thus, a signal may be considered 90\% significant if it has a false-alarm probability of 10\%.  However, our noise may not be precisely normal and independent due to potential unresolved pulsations and other systematic effects (such as atmospheric correlated noise) that may cause our data to be slightly correlated.  This may cause us to slightly overestimate our detection limits.  Our simulations of normal random data distributed with the same errors as our light curves have shown, however, that this effect is minimal.  We thus determine our detection limits by injecting sinusoidal signals, integrated over the observation's exposure times, at a single frequency and test all phases.  We then increase the amplitude of the signal until 90\% of all phases produce a peak in the periodogram at the injected frequency that is 90\% significant.  Finally, all frequencies accessible by the data are tested with this method.  A low frequency limit to the accessible frequencies is derived by testing the low frequency content of the comparison stars where we do {\it not} correct for color-airmass or other long period trends.  We report the low frequency limit as the highest frequency where the red-noise power drops below the 50\% false-alarm probability level.  All frequencies below this limit will be affected by the color-airmass and long period trend corrections and are thus not reliable.  For all observations, this low frequency limit is a few times larger than the sampling limit of the inverse of the observation length.  In all cases in this paper, the detection limits were largely independent of frequency allowing us to state one detection limit over a range of frequencies.  The results of our observations are shown in Table \ref{tbl:results} and summarized in the next section and Table \ref{tbl:candsum}.

\begin{deluxetable*}{llcccc}
\tabletypesize{\small}

\tablewidth{0pt}
\tablecaption{Results of Observations\label{tbl:results}}

\tablehead{
\colhead{Object} & \colhead{Date} & \colhead{Filter} & \colhead{Detection} & \colhead{Frequency} & \colhead{Figure} \\
	& & & \colhead{Limit (mmag)} & \colhead{Range (mHz)} &
 }
\tablecolumns{6}

\startdata

	SDSS~J0822+2743 &  2011/02/20 & $g^\prime$-SDSS & 8 & 0.5--11 & \ref{fig:j0822}  \\
	SDSS~J0849+0445 &  2011/01/17 & $g^\prime$-SDSS & 15 & 0.3--7 & \ref{fig:j0849} \\
	SDSS~J0917+4638 &  2011/02/20 & $g^\prime$-SDSS & 11 & 0.6--8 & \ref{fig:j0917} \\
	SDSS~J1330+0127 &  2007/05/30 & BG-39 & 44 & 0.3--5 & \ref{fig:j1330} \\
	SDSS~J1435+3733 &  2007/05/29 & BG-39 & 4 & 0.9--17 & \ref{fig:j1435} \\
	                                      &  2007/05/31 & BG-39 & 6 & 0.9--17 & \ref{fig:j1435} \\ 
	                                      &  2007/06/01 & BG-39 & 4 & 0.3--9 & \ref{fig:j1435} \\
	SDSS~J1448+0112 &  2007/05/31 & BG-39 & 34 & 0.5--7 & \ref{fig:j1448} \\
	SDSS~J2049+0005 &  2007/05/31 & BG-39 & 21 & 0.4--7 & \ref{fig:j2049} \\
	SDSS~J2240-0935 &  2006/08/04 & Gunn-$g$ & 22 & 0.3--7 & \ref{fig:j2240a} \\
	                                     & 2006/08/17 & Gunn-$g$ & 13 & 0.3--2.25 & \ref{fig:j2240a} \\
	                                     & 2006/09/19 & Gunn-$g$ & 14 & 0.6--7 & \ref{fig:j2240a} \\
	                                     & 2006/09/25 & Gunn-$g$ & 16 & 0.7--9 & \ref{fig:j2240a} \\
	                                     & 2006/12/05 & Gunn-$g$ & 48 & 0.5--9 & \ref{fig:j2240b} \\
	LP~400-22 &  2007/05/30 & $g^\prime$-SDSS & 4 & 0.5--15 & \ref{fig:lp400} \\
	NLTT~11748 &  2009/12/22 & $g^\prime$-SDSS & 5 & 0.2--7 & \ref{fig:nltt} \\
	                         &  2010/01/08 & $g^\prime$-SDSS & 4 & 0.2--7 & \ref{fig:nltt} \\
	PSR~J1012+5307 &  2010/03/30 & $g^\prime$-SDSS & 20 & 0.2--7 & \ref{fig:psrj1012} \\
	PSR~J1911-5958A &  2010/03/05 & F200LP & 16 & 0.05--3.25 & \ref{fig:psrj1012} \\
\enddata
\end{deluxetable*}

\section{ Summary of Observations and Results }
\label{sec:obssum}

In the course of this paper, several observing setups were employed.  Here we summarize the details of the telescopes and instrumentation used.

The William Hershel Telescope (WHT) is a 4.2-m telescope at the Roque de Los Muchachos Observatory on La Palma in the Canary Islands.  We used the ACAM imager and the SDSS-$g^\prime$ filter with a circular $8\farcm3$ diameter field-of-view in a 2$\times$2 pixel binning mode for an effective spatial resolution of $0\farcs5 \, {\rm pixel}^{-1}$.  Using the fast readout mode we maintain a total dead-time of $\approx$11\,s between exposures.  All observations using this setup were carried out using Service Time, generally at bright time (near the calendar full moon).

The Wisconsin Indiana Yale NOAO (WIYN) telescope is a 3.5-m telescope at the Kitt Peak National Observatory in Arizona.  We used the OPTIC imager \citep{how03.pasp115} and the broadband filter BG-39 ($\lambda_c\approx480\,{\rm nm},\,{\rm FWHM}\approx260$\,nm) with a $9\farcm5$$\times$$9\farcm5$ field-of-view in a 2$\times$2 pixel binning mode for an effective spatial resolution of $0\farcs28 \, {\rm pixel}^{-1}$.  The total dead-time between exposures was $\approx$10\,s.  All observations using this setup were carried out in a single observing run that occurred in late-May and early-June 2007.

The Palomar 60-inch (P60) telescope is a 1.5-m robotically controlled telescope at the Palomar Observatory on Mount Palomar in southern California.  We used the P60CCD imager and the Gunn-$g$ filter with a $12\farcm9$$\times$$12\farcm9$ field-of-view with a spatial resolution of $0\farcs38 \, {\rm pixel}^{-1}$.  We decreased readout times by reading out half of the two CCD chips for a total dead-time between exposures of $\approx$20--40\,s.  The large variation in dead-time was due to a software issue in the telescope software.  This setup was used in late 2006 and early 2007, however, most of the data taken during that time was for other projects or was too degraded by weather.

The Faulkes Telescope North (FTN) is a 2.0-m robotically controlled telescope on Haleakala, Hawaii and is part of LCOGT\footnote{http://www.lcogt.net}.  We used the Merope imager and the SDSS-$g^\prime$ filter with a $4\farcm75$$\times$$4\farcm75$ field-of-view in a 2$\times$2 pixel binning mode for an effective spatial resolution of $0\farcs28 \, {\rm pixel}^{-1}$.  The total dead-time between exposures was $\approx$22\,s.  This setup was used in late-December 2009 and early-January 2010.

We used the Wide Field Camera 3 (WFC3) UVIS channel and the F200LP filter ($\lambda_c \approx 490 \, {\rm nm}, \, {\rm FWHM} \approx 250$\,nm) on HST.  To decrease readout times and to avoid onboard memory buffer overflow we used the UVIS1-C512A-SUB subarray, a 512$\times$512 pixel sub-section of the CCD.  This gave us a $20\farcs5$$\times$$20\farcs5$ field-of-view and a total dead time of 55\,s.  This setup was used in March 2010 for 4 orbits (6 hours).

\begin{figure*}
	\centering
	\includegraphics{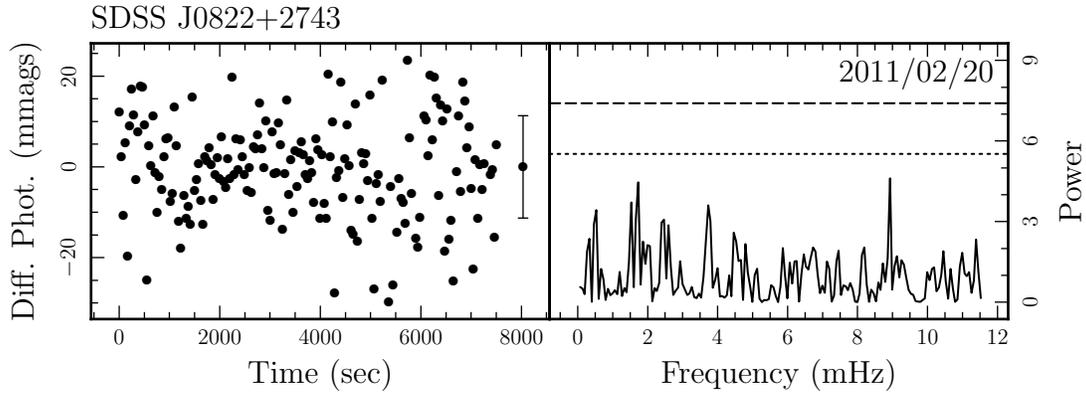}
	\caption{The light curve and Lomb-Scargle periodogram for SDSS~J0822+2743 for observations taken at WHT/ACAM on 2011 February 20.  The dashed (dotted) line in the periodogram represents the 10\% (50\%) false-alarm probability thresholds given the number of frequency bins.}
	\label{fig:j0822}
\end{figure*}

\begin{figure*}
	\centering
	\includegraphics{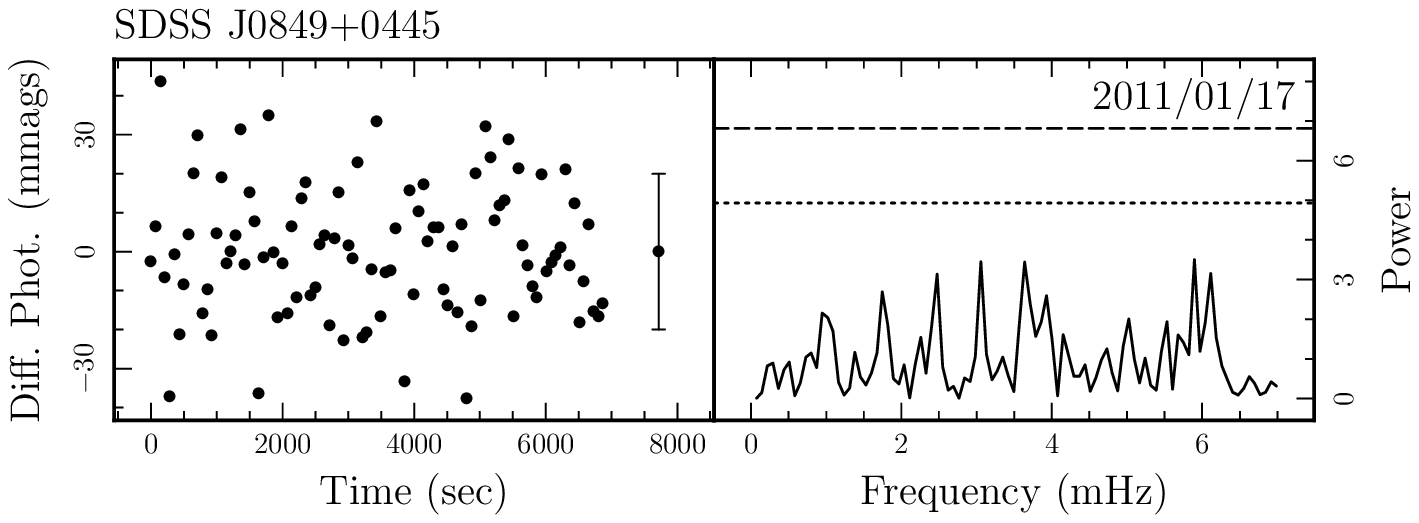}
	\caption{The light curve and Lomb-Scargle periodogram for SDSS~J0849+0445 for observations taken at WHT/ACAM on 2011 January 17.  The dashed (dotted) line in the periodogram represents the 10\% (50\%) false-alarm probability thresholds given the number of frequency bins.}
	\label{fig:j0849}
\end{figure*}

\begin{figure*}
	\centering
	\includegraphics{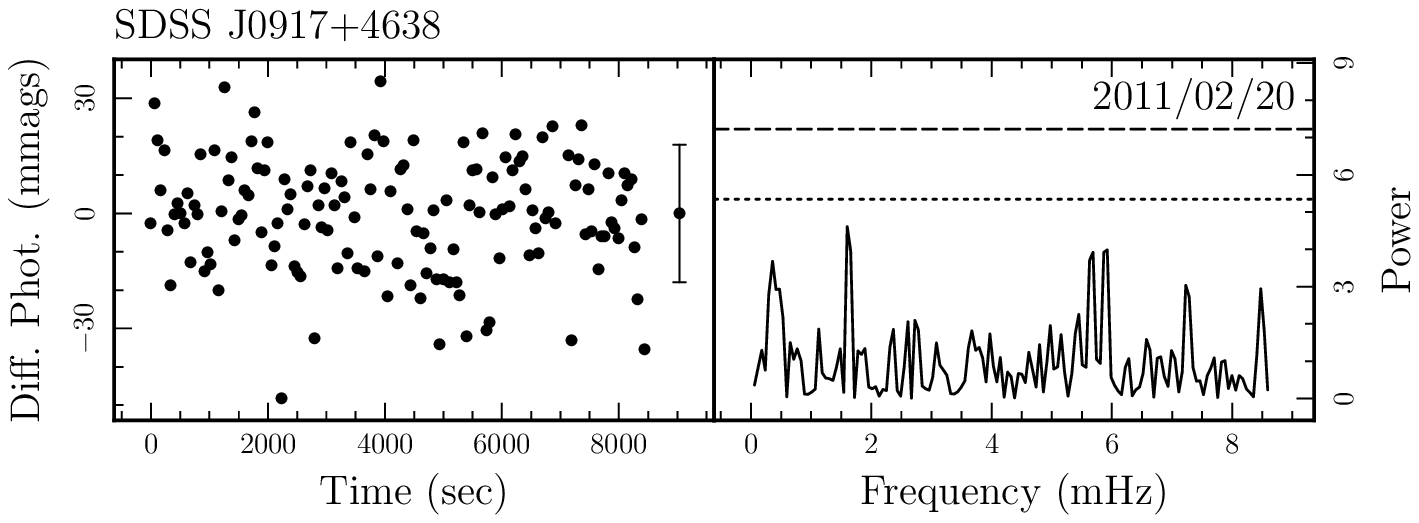}
	\caption{The light curve and Lomb-Scargle periodogram for SDSS~J0917+4638 for observations taken at WHT/ACAM on 2011 February 20.  The dashed (dotted) line in the periodogram represents the 10\% (50\%) false-alarm probability thresholds given the number of frequency bins.}
	\label{fig:j0917}
\end{figure*}

\begin{figure*}
	\centering
	\includegraphics{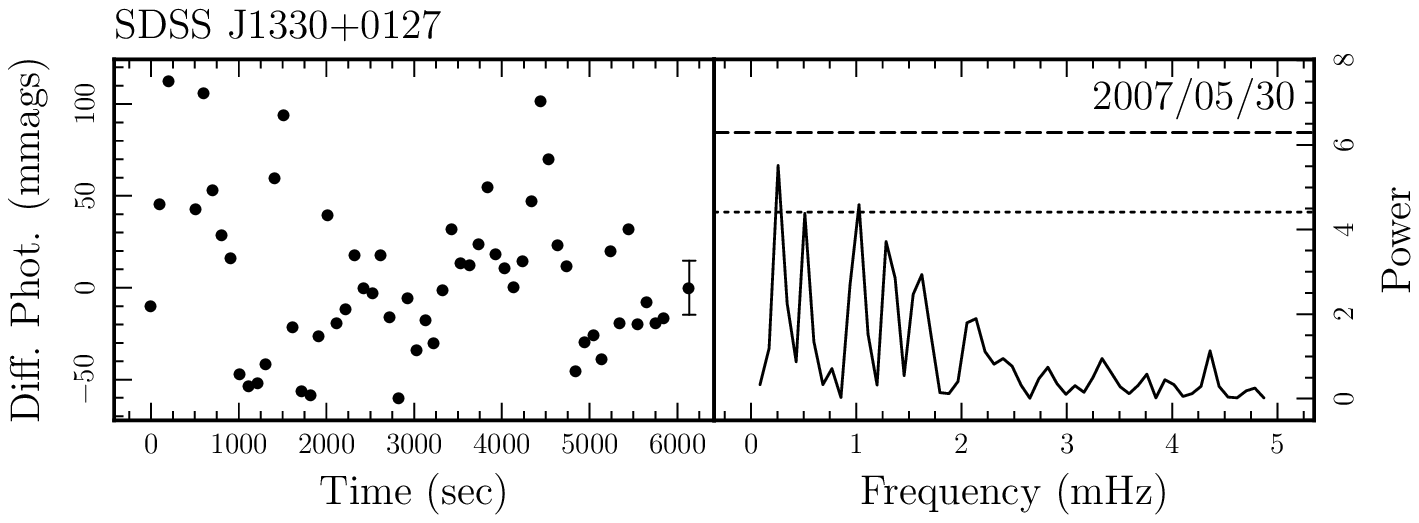}
	\caption{The light curve and Lomb-Scargle periodogram for SDSS~J1330+0127 for observations taken at WIYN/OPTIC on 2007 May 30.  The dashed (dotted) line in the periodogram represents the 10\% (50\%) false-alarm probability thresholds given the number of frequency bins.}
	\label{fig:j1330}
\end{figure*}

\begin{figure*}
	\centering
	\includegraphics{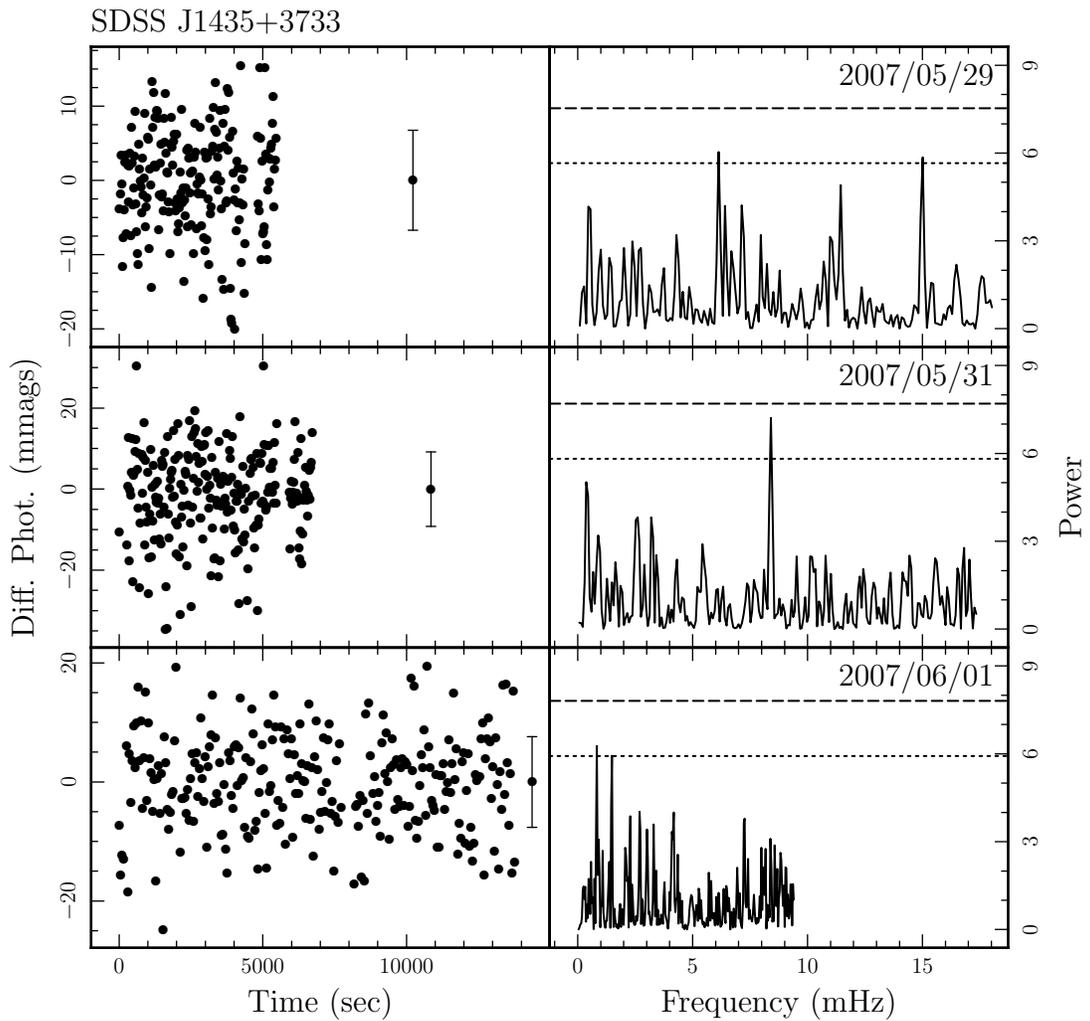}
	\caption{The light curves and Lomb-Scargle periodograms for SDSS~J1435+3733 for observations taken at WIYN/OPTIC on 2007 May 29, 31, and June 1.  The dashed (dotted) line in the periodogram represents the 10\% (50\%) false-alarm probability thresholds given the number of frequency bins.}
	\label{fig:j1435}
\end{figure*}

\begin{figure*}
	\centering
	\includegraphics{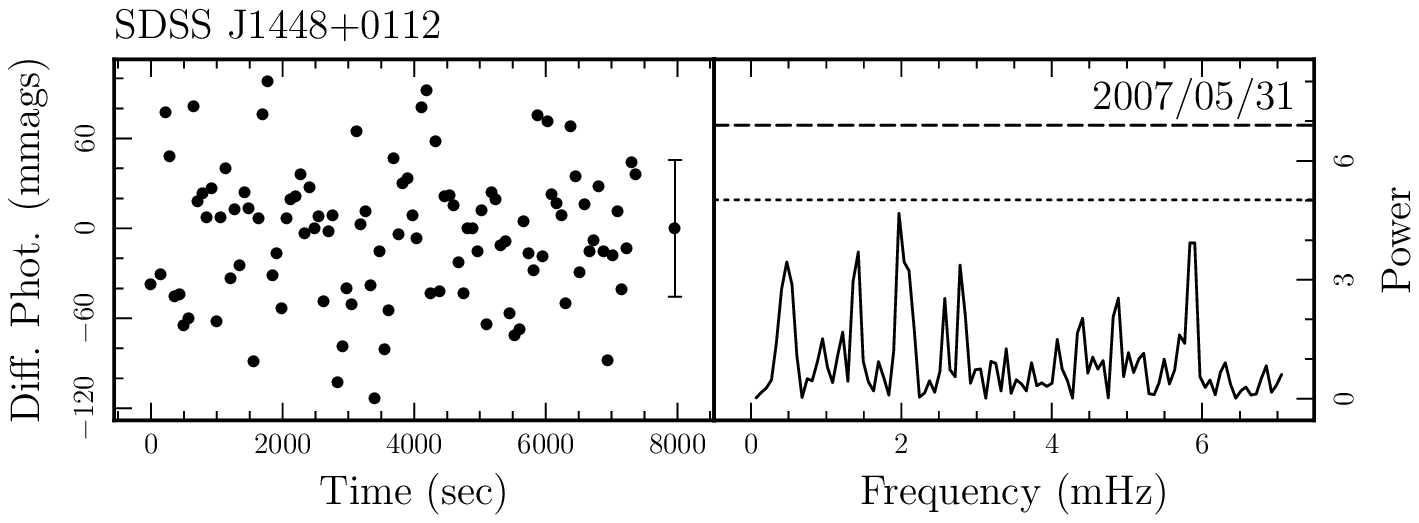}
	\caption{The light curve and Lomb-Scargle periodogram for SDSS~J1448+0112 for observations taken at WIYN/OPTIC on 2007 May 31.  The dashed (dotted) line in the periodogram represents the 10\% (50\%) false-alarm probability thresholds given the number of frequency bins.}
	\label{fig:j1448}
\end{figure*}

\begin{figure*}
	\centering
	\includegraphics{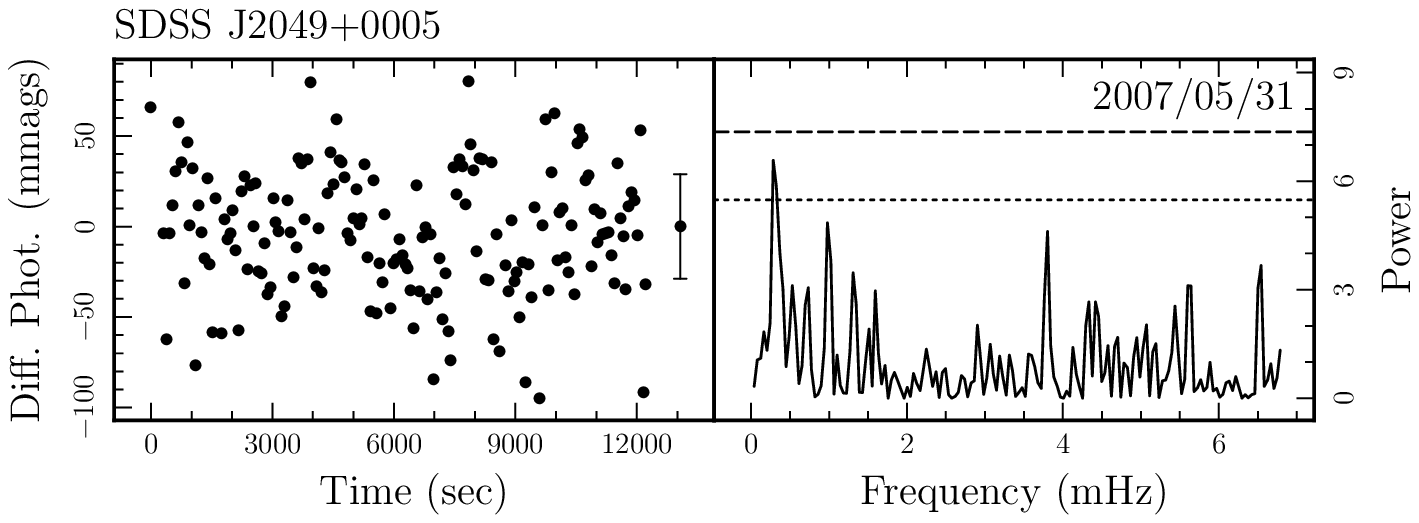}
	\caption{The light curve and Lomb-Scargle periodogram for SDSS~J2049+0005 for observations taken at WIYN/OPTIC on 2007 May 31.  The dashed (dotted) line in the periodogram represents the 10\% (50\%) false-alarm probability thresholds given the number of frequency bins.}
	\label{fig:j2049}
\end{figure*}

\begin{figure*}
	\centering
	\includegraphics{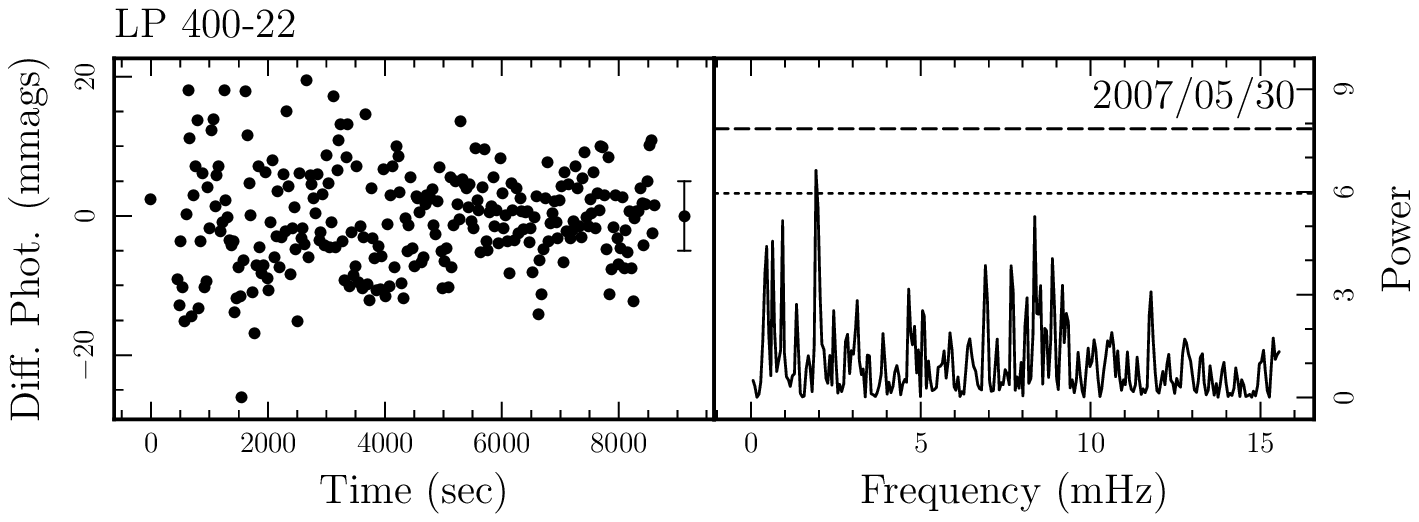}
	\caption{The light curve and Lomb-Scargle periodogram for LP~400-22 for observations taken at WIYN/OPTIC on 2007 May 30.  The dashed (dotted) line in the periodogram represents the 10\% (50\%) false-alarm probability thresholds given the number of frequency bins.}
	\label{fig:lp400}
\end{figure*}

\begin{figure*}
	\centering
	\includegraphics{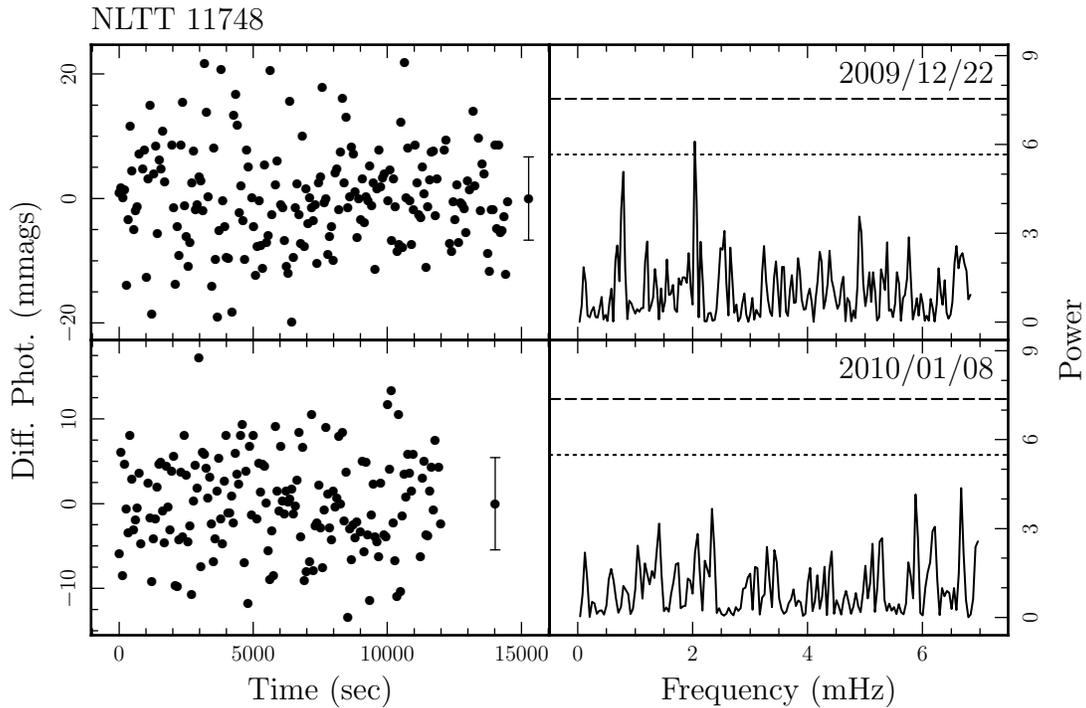}
	\caption{The light curves and Lomb-Scargle periodograms for NLTT~11748 for observations taken at FTN/Merope on 2009 December 22 and 2010 January 8.  The dashed (dotted) line in the periodogram represents the 10\% (50\%) false-alarm probability thresholds given the number of frequency bins.}
	\label{fig:nltt}
\end{figure*}

\begin{figure*}
	\centering
	\includegraphics{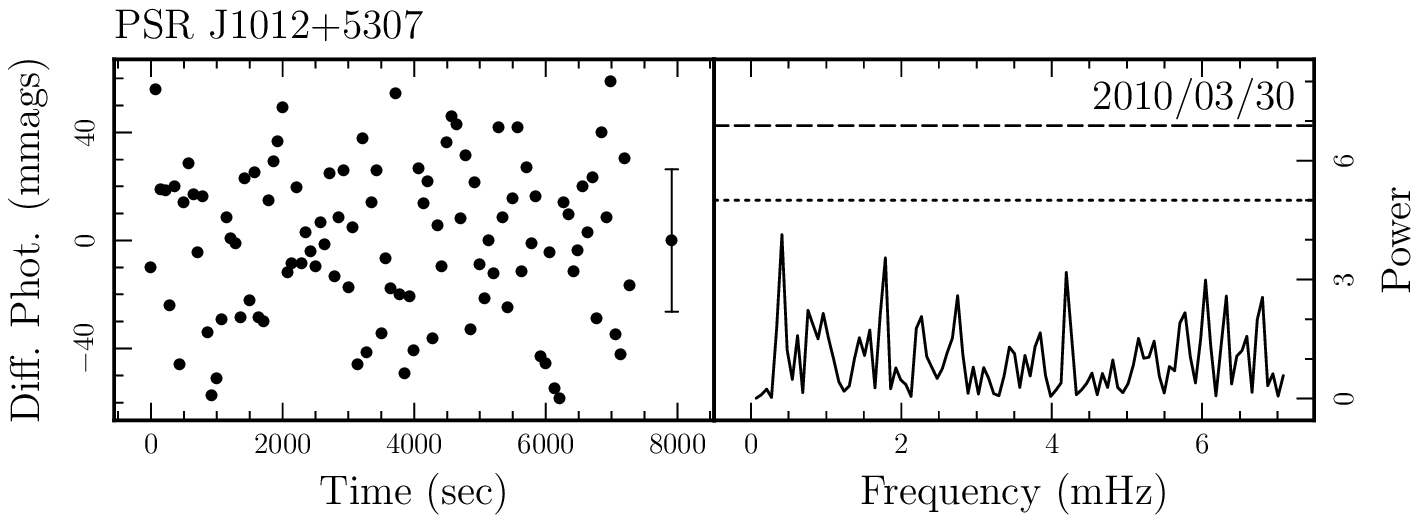}
	\caption{The light curve and Lomb-Scargle periodogram for PSR~J1012+5307 for observations taken at WHT/ACAM on 2010 March 30.  The dashed (dotted) line in the periodogram represents the 10\% (50\%) false-alarm probability thresholds given the number of frequency bins.}
	\label{fig:psrj1012}
\end{figure*}

\begin{figure*}
	\centering
	\includegraphics{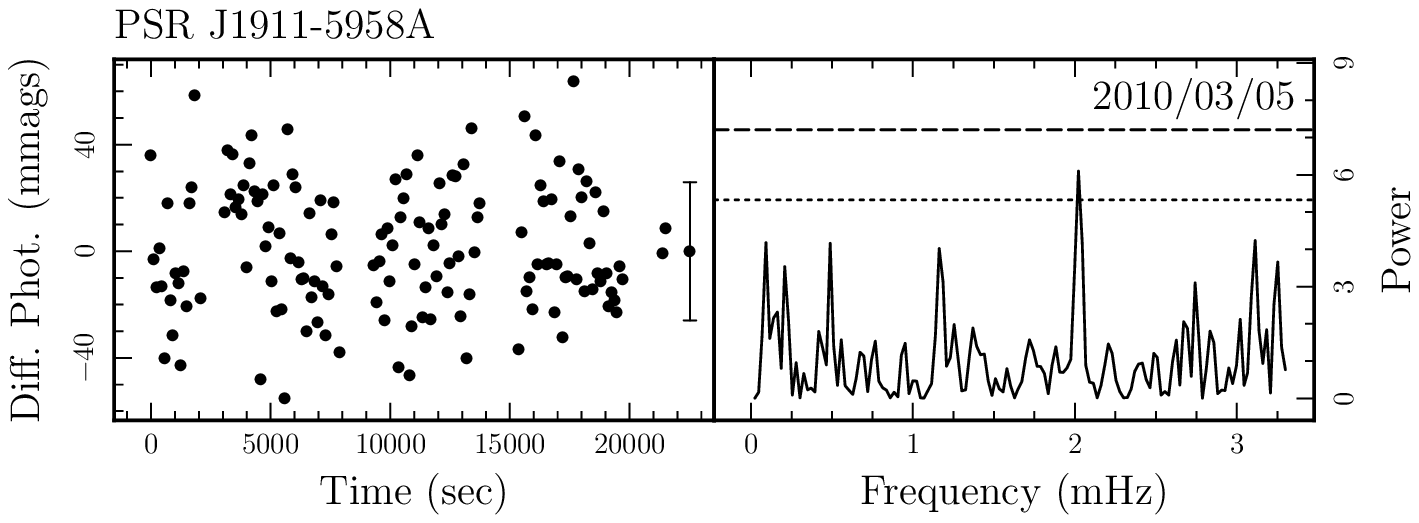}
	\caption{The light curve and Lomb-Scargle periodogram for PSR~J1911-5958A for observations taken at HST/WFC3 on 2010 March 3.  The dashed (dotted) line in the periodogram represents the 10\% (50\%) false-alarm probability thresholds given the number of frequency bins.}
	\label{fig:psrj1911}
\end{figure*}

Most of our observations and objects were routine and the results are well summarized in Tables \ref{tbl:candsum}, \ref{tbl:detobs}, and \ref{tbl:results} and Figures \ref{fig:j0822}--\ref{fig:psrj1911}.  However, a few of our objects and observations require special discussion.

{\bf SDSS~J1330+0127} was observed once using the WIYN/OPTIC setup.  No pulsations were detected to a limit of 44\,mmags over a frequency range of 0.3--5\,mHz.  This observation was plagued by intermittent cirrus clouds and a near full moon creating a high sky background.  This object has a measured $\log g = 5.56 \pm0.16$ and $T_{\rm eff} = 8,091 \pm 75$\,K \citep{eis06.apjs167}.  The SDSS spectra were likely contaminated by a companion star's flux.  A re-analysis by \citet{sil06.aj131} measured $\log g = 8.1 \pm 0.2$ and $T_{\rm eff} = 10,617 \pm 452$\,K.  This object is likely not a He WD.

{\bf SDSS~J1435+3733} is a partially eclipsing binary system of a C/O WD and M-dwarf discovered by \citet{ste08.apj677} using the FTN/Merope setup.  All frames in eclipse were removed for pulsation analysis.  No pulsations were detected to a limit of 4\,mmags over a frequency range of 0.9--17\,mHz.  This object has a measured $\log g = 6.86 \pm 0.04$ and $T_{\rm eff} = 11,062 \pm 58$\,K \citep{eis06.apjs167}.  The SDSS spectra were definitely contaminated by a companion star's flux \citep{ste08.apj677,reb07.mnras382,pyr09.mnras394}.  A re-analysis by \citet{reb07.mnras382} and \citet{pyr09.mnras394} measured $\log g = 7.62 \pm 0.12$ and $T_{\rm eff} = 12,536 \pm 488$\,K which gives a derived model mass of 0.41$M_{\odot}$, near the C/O WD and He WD composition boundary.

{\bf SDSS~J2049+0005} was observed once using the WIYN/OPTIC setup.  No pulsations were detected to a limit of 21\,mmags over a frequency range of 0.4--7\,mHz.  This observation was plagued by intermittent cirrus clouds and a near full moon creating a high sky background.  This object has a measured $\log g = 5.48 \pm 0.10$ and $T_{\rm eff} = 8,660 \pm 144$\,K \citep{eis06.apjs167}.  \citet{eis06.apjs167} noted multiple minima in their model fits as well as discrepant SDSS-$g^\prime$ band flux compared to best fit models and suggested a $\log g < 5.0$ was likely.  \citet{kil07.apj660} confirmed this finding and conclude that this star is likely a distant A2 star.

{\bf SDSS~J2240-0935} was observed four times using the P60/P60CCD setup.  No pulsations were detected to a limit of 14\,mmags over 0.6--7\,mHz.  This object has a measured $\log g = 6.96 \pm 0.09$ and $T_{\rm eff} = 11,449 \pm 205$\,K \citep{eis06.apjs167}.  The SDSS spectra were likely contaminated by a companion star's flux.  A re-analysis by \citet{hel09.aap496} measured $\log g = 7.5$ and $T_{\rm eff} = 13,000$\,K.

{\bf NLTT~11748} is the first totally eclipsing detached double white dwarf binary system discovered by \citet{ste10.apj716} using the FTN/Merope setup.  All frames in eclipse were removed for pulsation analysis.  No pulsations were detected to a limit of 4\,mmags over 0.2--7\,mHz.  This object has a measured $\log g = 6.54 \pm 0.05$ and $T_{\rm eff} = 8,690 \pm 140$\,K \citep{kil10.apj721} with a derived mass of 0.18$M_{\odot}$.  \citet{shp10.apj725} found direct evidence for Doppler beaming in a long timescale photometric survey.  The Doppler beaming signal was removed in the low order polynomial division in the reduction and should only affect frequencies less than 0.1\,mHz (half the orbital period of 5.6\,hours).  In the future, eclipse and radial velocity analysis will produce a high precision measurement of the masses and radii of both components in this object.  Since its discovery, three additional eclipsing He WD binary systems have been discovered \citep{pyr11.arx,bro11.apj737,par11.apj735}.

{\bf PSR~J1012+5307} was observed one time using the WHT/ACAM setup.  No pulsations were detected to a limit of 20\,mmags over the frequency range of 0.2-7\,mHz.  This object has a measured $\log g = 6.34 \pm 0.20$ and $T_{\rm eff} = 8,670 \pm 300$\,K with a derived model mass of 0.18$M_{\odot}$ \citep{cal98.mnras298} and also $\log g = 6.75 \pm 0.07$ and $T_{\rm eff} = 8,550 \pm 25$\,K with a derived model mass of 0.21$M_{\odot}$ \citep{van96.apj467}.  Neither of these measurements is obviously superior.

{\bf PSR~J1911-5958A} was observed one time using the HST/WFC3 setup over 4 orbits.  The photometric analysis for this object was uniquely different than all other objects.  Due to crowding in the field (PSR~J1911-5958A is in a globular cluster) as well as the unique PSF of HST, aperture photometry is not the most precise method.  PSF-fitting photometry is ideal since the PSF of the WFC3 instrument is well understood (Jay Anderson, private communication).  We used the PSF-fitting software developed by Jay Anderson (similar to \citealt{and06.tech06}) to extract the photometry from our data.  Since most atmospheric and sky effects are nonexistent in space, variable aperture photometry was not required or used.  We used 36 comparison stars with magnitudes between 17.5--25.0 to ensure the quality of our photometry.  Due to a stray light artifact, our background was increased by $\approx$30\%.  No pulsations were detected to a limit of 16\,mmags over the frequency range of 0.05-3.25\,mHz.  This object has a measured $\log g = 6.44 \pm 0.20$ and $T_{\rm eff} = 10,090 \pm 150$\,K with a derived model mass of 0.18$M_{\odot}$ \citep{bas06.aap456}.

\section{ Conclusions }
\label{sec:conc}

Of the 12 objects we observed in this paper, seven are likely very low mass ($<$0.2\,$M_{\odot}$) He WDs.  Whether or not these observations offer strong constraints on theory is currently an open matter.  First we must consider the accuracy of the $\log g$ and $T_{\rm eff}$ measurements that place these objects within the predicted pulsation parameter space.  Unfortunately, very low mass WD atmospheres are not well understood.  Due to their low surface temperatures ($T_{\rm eff}\lesssim9$,000\,K), the H atmospheres are not sufficiently ionized and neutral broadening of the H absorption lines plays an important role.  The theory of neutral broadening is not well understood and may account for significant errors in line widths (\citealt{bar00.aap363,all04.aap424,tre10.apj712}; private communication, Detlev Koester).  As such, the reported errors on $\log g$ and $T_{\rm eff}$ for these objects are always the statistical error of the fit to theoretical atmosphere grids and do not take into account the uncertainties of the theory itself which can increase the uncertainty by as much as a factor of two or more (private communication, Detlev Koester).  Further, the uncertainty of the atmosphere theory is likely not random but rather a shift in a specific direction in the $T_{\rm eff}$--$\log g$ plane, affecting the parameter space location of all very low mass He WDs.  This is why NLTT~11748 is such an important eclipsing binary system \citep{ste10.apj716}.  As a potential double-lined spectroscopic eclipsing binary, it could offer high precision \textit{model independent} determinations of both the mass and radius of the He WD component providing a high precision constraint on its gravity for which any atmospheric gravity measurement must match.

Of added concern is the location of the instability strip ``blue" edge itself.  \citet{ste10.apj718} derived the ``blue" edge to be where pulsation modes of eigenvalues $\ell=1$ and $n=1$ were driven.  This offers a temperature upper limit to the $\ell=1$ instability strip because the ``blue" edge shifts redward when higher radial orders of $n$ are considered.  In the few cases where pulsation mode eigenvalues have been identified observationally (e.g. HL~Tau~76, \citealt{pec06.aap446}; G117-B15A and R548, \citealt{bra98.apjs116}) the highest amplitude observed modes range in $n$ from 1 to several 10s.  If very low mass He WDs preferentially only excite the higher radial order modes, then the blue edge could shift redward by a few 100\,K potentially removing our three best candidates from the potential instability region. Uncertainties in the precise convective efficiency and the lack of a detailed non-adiabatic analysis could further shift this blue edge by another several 100\,K in either direction \citep{win82.apj262,fon08.pasp120}.

An additional observational concern is whether our detection limits are truly meaningful.  Only NLTT~11748, LP~400-20, SDSS~J0822+2743, and SDSS~J1435+3733 met our fiducial pulsation limit of 10\,mmags and only NLTT~11748, SDSS~J1435+3733, and SDSS~J2240-0935 were observed more than once.  It is well known that pulsation amplitudes can be small, 3--4\,mmag, and vary in  detectability from one night to the next \citep{muk06.apj640,ste08.pasp120}.  Therefore additional observations are required for many of our objects to both tighten their detection limits and gain multiple epochs of observation so that these null detections can be even more convincing.

Finally, the theoretical analysis by \citet{ste10.apj718} involved only an extension of the Brickhill convective driving criterion using adiabatic assumptions which may not be valid at the lower gravities of the very low mass He WDs.  Further, very little theoretical work has been done between 0.2-0.45\,$M_{\odot}$ \citep{alt04.mnras347,arr06.apj643}.  To address these concerns, an extension of the non-adiabatic driving work by \citet{bri91.mnras251} and \citet{wu99.apj519} should be used to firmly predict the location of the He WD instability strip.

\acknowledgments

We thank both referees for comments that improved our paper.  This work was supported by the National Science Foundation under grants PHY 05-51164 and AST 07-07633.  DLK was partially supported by NASA through Hubble Fellowship Grant \#01207.01-A awarded by the STScI which is operated by AURA, Inc., for NASA, under contract NAS 5-26555.  This paper uses observations obtained with facilities of the Las Cumbres Observatory Global Telescope.  Support for Program number HST-GO-11581.01-A was provided by NASA through a grant from the Space Telescope Science Institute, which is operated by the Association of Universities for Research in Astronomy, Incorporated, under NASA contract NAS5-26555.

\begin{figure*}
	\centering
	\includegraphics{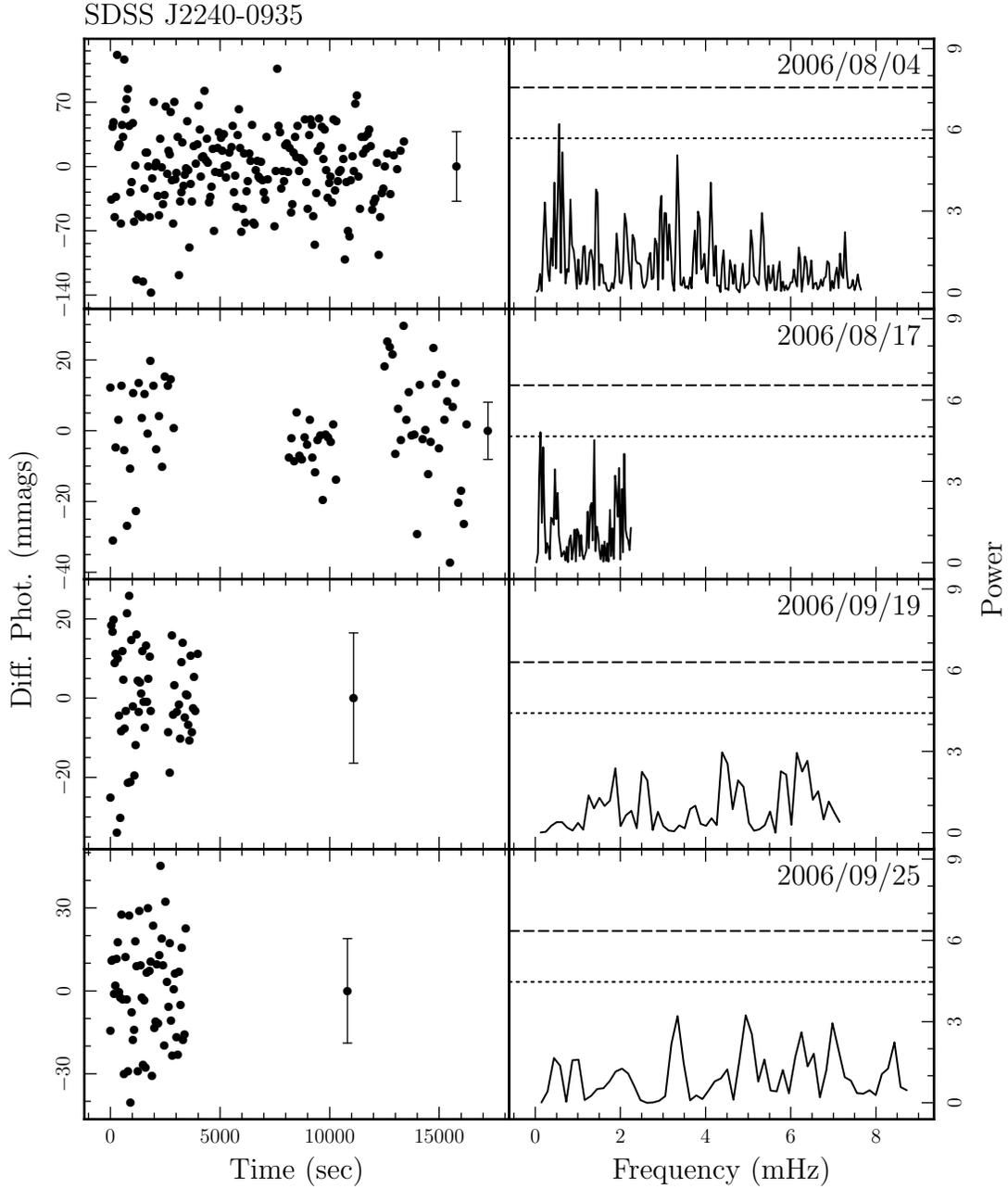}
	\caption{The light curves and Lomb-Scargle periodograms for SDSS~J2240-0935 for observations taken at P60/P60CCD on 2006 August 4 and 17 and September 19 and 25.  The dashed (dotted) line in the periodogram represents the 10\% (50\%) false-alarm probability thresholds given the number of frequency bins.}
	\label{fig:j2240a}
\end{figure*}

\begin{figure*}
	\centering
	\includegraphics{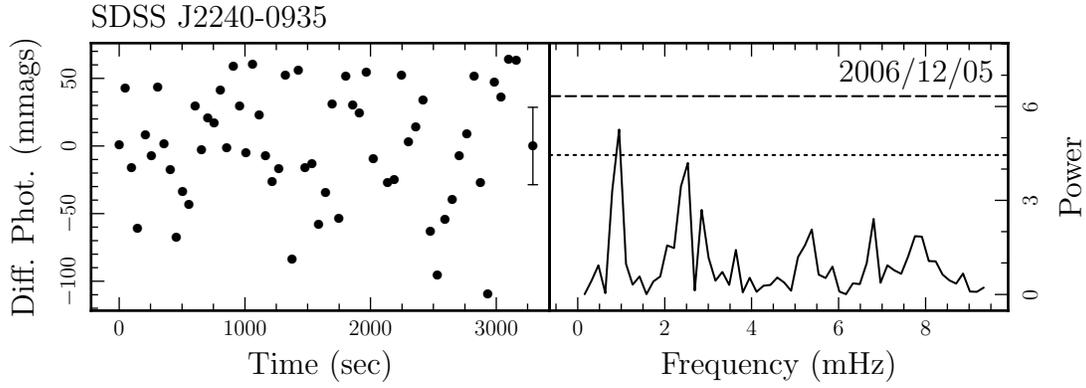}
	\caption{Continuation of Figure \ref{fig:j2240a} for observations taken on 2006 December 5.}
	\label{fig:j2240b}
\end{figure*}

\end{document}